\documentclass[12pt,preprint]{aastex}
\usepackage{epsfig}
\shorttitle{Oxygen abundance of open cluster dwarfs}
\shortauthors{Shen, Liu, Zhang, Jones \& Lin}

\begin{document}
\title{ Oxygen abundance of open cluster dwarfs}

\author{Z.-X. Shen}
\affil{Department of Astronomy, Peking University, Beijing 100871,
        P. R. China;
{shenzx@bac.pku.edu.cn}}
\and

\author{X.-W. Liu}
\affil{Department of Astronomy, Peking University, Beijing 100871,
        P. R. China; liuxw@bac.pku.edu.cn}
\and

\author{H.-W. Zhang}
\affil{Department of Astronomy, Peking University, Beijing 100871,
        P. R. China;}
\and

\author{B. Jones}
\affil{ UCO/Lick Observatory, Departments of Astronomy and Astrophysics, University
    of California, Santa Cruz CA 95064, U.S.A.}
\and

\author{D. N. C. Lin}
\affil{ UCO/Lick Observatory, Departments of Astronomy and Astrophysics, University
    of California, Santa Cruz CA 95064, U.S.A.}

\begin{abstract} We present oxygen abundances of dwarfs in the young open
cluster IC\,4665 deduced from the \ion{O}{1} $\lambda$7774 triplet lines and of
dwarfs in the open cluster Pleiades derived from the [\ion{O}{1}] $\lambda$6300
forbidden line.  Stellar parameters and oxygen abundances were derived using
the spectroscopic synthesis tool SME (Spectroscopy Made Easy).  We find a
dramatic increase in the upper boundary of the \ion{O}{1} triplet abundances
with decreasing temperature in the dwarfs of IC\,4665, consistent with the
trend found by Schuler et al. in the open clusters Pleiades and M\,34, and to a
less extent in the cool dwarfs of Hyades (Schuler et al. 2006a) and UMa (King
\& Schuler 2005).  By contrast, oxygen abundances derived from the [\ion{O}{1}]
$\lambda$6300 forbidden line for stars in Pleiades and Hyades (Schuler et al.
2006b) are constant within the errors.  Possible mechanisms that may lead a
varying oxygen triplet line abundance are examined, including systematic errors
in the stellar parameter determinations, the NLTE effects, surface activities
and granulation.  The age-related effects stellar surface activities
(especially the chromospheric activities) are suggested by our analysis to
blame for the large spreads of oxygen triplet line abundances.

\end{abstract}


\section{Introduction} 

Oxygen is an abundant element of particular relevance for the Galactic chemical
evolution and its formation history. Oxygen in long-lived low-mass stars
represents the chemical composition of the gas at the time those stars are
formed and carries information of the chemical history of stars in different
populations in our Galaxy. It is a bona fide primary element formed exclusively
in the interiors of massive stars and then released to the interstellar medium
(ISM) via Type\,II supernova explosions. By analyzing oxygen abundances
relative to the iron group elements in the atmosphere of different type stars,
one can trace the history of SN\,II feedback and the occurrence rates of
Type\,II and Type\,Ia supernova explosions. 

Recent reviews on stellar oxygen abundance analyses are given by King (1993),
Israelian et al. (1998), Nissen et al. (2002) and Takeda (2003).  Oxygen has a
limited number of lines in the visual part of stellar spectra. Apart from
molecular OH lines in the ultraviolet (UV) and in the infrared (IR), oxygen
abundances are traditionally determined from the high-excitation \ion{O}{1}
3s\,$^5$S$^{\rm o}$ -- 3p\,$^5$P $\lambda\lambda$7771.94, 7774.16, 7775.39
triplet permitted lines or from the much weaker [\ion{O}{1}] 2p$^4$\,$^3$P --
2p$^4$\,$^1$D $\lambda\lambda$6300.30, 6363.78 forbidden lines.  Both methods
have their advantages and disadvantages. The triplet lines are strong and
therefore easy to measure, and are free from blending effects.  However, their
strengths can be significantly affected by effects such as deviations from the
local thermal equilibrium (LTE) (e.g., Kiselman 1993). On the other hand,
although the $\lambda\lambda$6300,6363 forbidden lines are believed to be free
from the non-LTE (NLTE) effects (e.g., Kiselman 1991; Takeda 2003), they are
much weaker than the triplet lines and blended with lines from other species.
The stronger component $\lambda$6300 of the doublet is blended with a
\ion{Ni}{1} line at 6300.34\,\AA\,(Lambert 1978; Johansson et al. 2003),
whereas the weaker $\lambda$6363 line can be contaminated by a weak CN red
system line (Lambert 1978). Discrepancies between oxygen abundances deduced
from the two indicators for field stars have long been observed and the possible causes have
been much discussed.  In particular, in metal-poor stars, a distinct trend has
been reported in the sense that the permitted lines tend to yield
systematically higher oxygen abundances than the forbidden lines (e.g., Cavallo
et al. 1997).  An inspection of the parameter-dependence of the discordance
indicates that the extent of the discrepancy tends to be comparatively lessened
for stars of high $T_{\rm eff}$ and/or ${\rm log}\,g$ (Takeda 2003).

Many efforts have been attempted to resolve the discrepancy.  King (1993)
suggested that the temperature scale for metal-poor dwarfs is probably
150-200\,K too low and revised the scale upward to bring the oxygen abundances
derived from the two indicators into better agreement.  However, the
assumptions made by King (1993) in determining the theoretical colors were
 challenged by Balachandran \& Carney (1996).  A controversial factor
in oxygen abundance determinations has been the role of the NLTE corrections.
Kiselman \& Nordlund (1995) re-examined the treatment of the NLTE effects in
earlier studies and concluded that the cross sections for inelastic collisions
with neutral hydrogen may have been overestimated.  They argued that while
adjusting stellar parameters cannot resolve the oxygen abundance discrepancy,
it can be by resorting to more realistic 3D models.  NLTE corrections of oxygen
abundances deduced from the triplet lines have been performed by many
investigators (e.g., Tomkin et al. 1992; Mishenina et al.  2000; Carretta et
al. 2000; Nissen et al. 2002; Takeda 2003).  It is shown that the formation of
the \ion{O}{1} triplet is quite simple in the sense that only the transition
related quantities are important, but not the details of the atomic model. It
is also found that the NLTE effects of those particular lines can be well
described by a classic two-level-atom model and that the metallicity cannot be
an essential factor (c.f. Takeda 2003 and the references therein).   

While the physical cause of the discrepancy between the oxygen abundances
derived from the two types of line remains an open question, a modest view is
that the abundances from the forbidden lines are probably less problematic
given their weak $T_{\rm eff}$-sensitivity and negligible NLTE effects, and should
therefore be more reliable than those derived from the permitted lines.  On the
other hand, given the weakness of the forbidden lines, the permitted lines
remain indispensable, in particular for the analysis of warm stars and for
stars of low metallicities.

Members of an open cluster are assumed to form from a chemically homogeneous
cloud in a short time scale, such that they should all have the same chemical
composition. Stars in a cluster are distinguished only by mass and thus provide
good test beds to probe the underlying physical cause of the discrepant oxygen
abundances determined from the permitted lines and from the forbidden lines.
Schuler et al.  (2004) reported their analysis of oxygen abundances for a
sample of late F, G and K dwarfs of the open clusters Pleiades (100 Myr) and
M\,34 (250 Myr). They find a dramatic increase of the \ion{O}{1} triplet
abundance with decreasing effective temperature in both clusters. Later on,
similar but to a less extent trends are found in UMa (King \& Schuler 2005) and
Hyades (Schuler et al. 2006a; for stars with $T_{\rm eff} \la$ 6000\,K).  In
Hyades, oxygen triplet abundances of stars with $T_{\rm eff} \ga$ 6000\,K
increase with increasing temperature due to NLTE effect.  The phenomenon were
not found in the other open clusters probably because of their limited star
samples.  In contrast to the trend in oxygen triplet line abundances,
[\ion{O}{1}] $\lambda$6300 line abundances of three Pleiades stars (Schuler et
al. 2004) and 8 Hyades stars (Schuler et al. 2006b)
are nearly constant and yield average values much lower compared to the triplet
line results. Schuler et al. (2004, 2006a) suggest that surface inhomogeneities
rather than chromospheric activities are possibly the main cause of the
anomalous oxygen triplet line abundances. On the other hand, Morel \& Micela
(2004) compare oxygen triplet line and forbidden line abundances for stars
spanning a wide range of activity level and find that the magnitude of the
abundance discrepancy increases with increasing level of chromospheric/coronal
activities. 

Here we present oxygen triplet line abundances for a sample of dwarfs
of the open cluster IC\,4665 (35 Myr, Mermilliod 1981). In our
previous analysis (Shen et al. 2005), we showed that within the
measurement uncertainties the iron abundance is uniform, with a
standard deviation of 0.04 dex. This upper limit in the dispersion of
[Fe/H] among the IC 4665 member stars was used to infer that the total
reservoir of heavy elements retained by the nascent disks is limited.
Nevertheless, gas giant planets can form through core-accretion in
protostellar disks around these stars near snow lines where the
surface density of water ice may be significantly enhanced (Ida \& Lin
2004, Ciesla \& Cuzzi 2006).  In this case, the emergence of gas giants
may introduce a dispersion in [O/H] among the cluster member stars.

A first step in this difficult task is to calibrate the Oxygen
abundance as a function of stellar parameters.  In addition to our
sample of the IC 4665 member stars, forbidden line oxygen abundances
for several Pleiades dwarfs are also obtained. Section~2 describes the
observations and procedures of data reduction. Oxygen abundances
determined from the two types of line are presented in Section~3.
Possible explanations for the abundance determination discrepancy are
discussed in Section~4, followed by a brief summary in Section~5.

\section{Observations and data reduction}

Observations of both clusters were carried out in October 1999 and October 2000
using the HiRes spectrograph (Vogt 1992)
mounted on the Keck I 10\,m telescope. The spectra were recorded on
a Tektronix $2048\times 2048$ CCD of $24\times 24$\,$\mu{\rm m}$ pixel size.
For stars of IC\,4665, the spectra spanned a wavelength range from $\sim 6300$
to 8730\,\AA, split into 16 orders, with small interorder gaps amongst
them. The integration time ranged from 10 minutes to half an hour, yielding
signal-to-noise ratios ranging from $\sim 30$ to 150 per resolution element.
For Pleiades stars, the spectra covered from $\sim 4500$ to 6900\,\AA\,
with signal-to-noise ratios of $\sim$ 200 to 400.  The Pleiades spectra have
previously been used by Wilden et al. (2002) to study the metallicity
dispersion amongst the member stars. A detailed description of those spectra
can be found there.  All spectra have a resolving power of about 60,000.

All spectra were reduced using {\sc iraf} in a similar manner as described in
Soderblom et al. (1993).  The noao.imred.echelle package was used for
flat-fielding, scattered light removal, order extraction and wavelength
calibration. The latter was achieved using exposures of a Th-Ar lamp. 
 
We did not include all 18 IC\,4665 stars which have been analysis in our
first paper devoted to IC\,4665
(Shen et al. 2005; Paper~I thereafter). The three cool stars P\,332, P\,349 \& P\,352 are found to have
abnormal stronger ${\rm H}\alpha$ absorption than that of normal dwarfs at the
same temperatures. Given that the membership of the three stars are only based
on the color-magnitude diagram, it is likely to us that they are
background subgiants but not members in IC\,4665. We thus exclude them from our
sample so that they would not affect our oxygen abundance analysis results.
Further proper motion detection is need to clarify the membership of them.    
The existence of these subgiants in our sample would not affect our conclusions in Paper~I.
 
By performing on our spectra the cross-correlation analysis with solar
spectrum, we found that P\,19, our hottest star is a binary. The separation of
the two members on the spectrum is 1.5\,\AA. Based on the relative flux at
${\rm H}\alpha$, we suggest that the companion is a hot A type star. Given that
A stars have much weaker flux than the late F stars at the red part of the
spectrum and the spectrum line separation of them are large, the existence of a
companion would not affect our analysis results either on stellar parameters or
on element abundances. P\,19 is thus kept in our sample.

\section{Abundance determinations}

Spectral synthesis analyses were carried out using the software package SME
(Spectroscopy Made Easy) developed by Valenti \& Piskonov (1996).  SME can be
used to determine stellar and atomic parameters by matching the synthetic
spectrum to the observed one. It consists of a spectral synthesis code written
in C++ and a parameter optimization routine written in IDL. It uses Kurucz
stellar atmospheric models. The input parameters include $T_{\rm eff}$, ${\rm
log}\,g$, radial and rotational velocities, micro- and macro-turbulence
velocities, element abundances, and a list of spectral line atomic data (${\rm
log}\,gf$ and the van der Waals damping constants).  The overall
metallicity, quantified by parameter [M/H], is used to interpolate a grid of
model atmospheres and to scale the solar abundance pattern (except for helium
and the elements with individually solved abundances) when calculating the
opacities.  [M/H] is an independent model parameter, rather than a quantity
constructed from the abundances of individual elements. Effects on stellar
atmosphere caused by deviations of individual element abundances from the solar
abundance pattern are neglected. This is appropriate as the current work only
deals with stars of approximately solar metallicity. We assume the [M/H] of 
stars in IC\,4665 and Pleiades to be zero. SME solves the radiative transfer to generate
a synthetic spectrum. A nonlinear least squares algorithm is then used to solve
for any subset of the aforementioned input parameters. The radiative transfer
routine assumes LTE, no opacity from molecular lines and negligible magnetic
field. 

\begin{figure}
\plotone{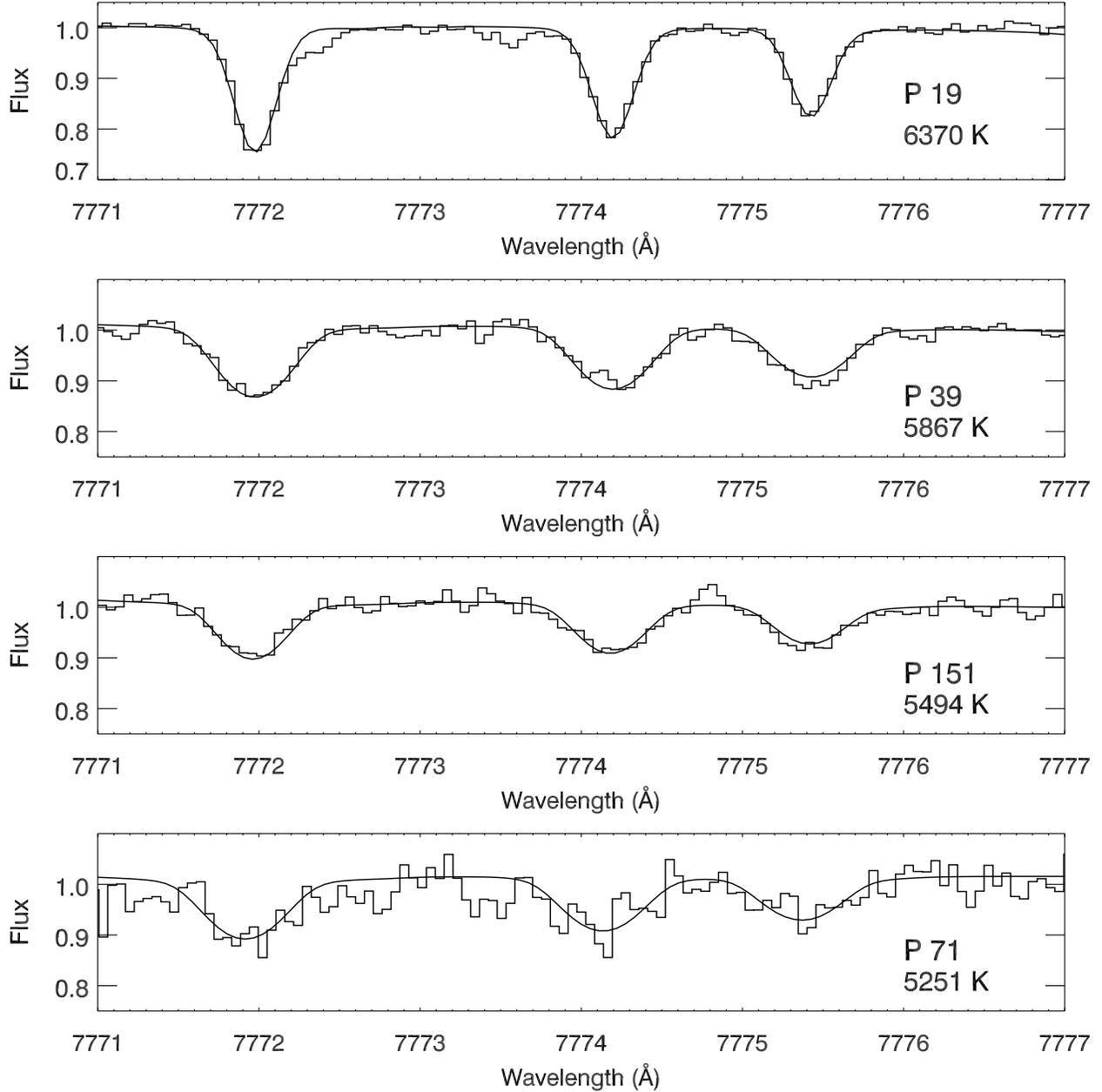}
\caption{Sample spectra of four IC\,4665 stars centered on the \ion{O}{1}
triplet lines. The spectra were selected to illustrate the full range of S/N
ratios achieved amongst the sample stars. Also plotted are the best fit
synthetic spectra (smooth curves) obtained for individual stars.}
\label{otrip}
\end{figure}

A 10\,\AA\ wide spectral segment centered at 7774\,\AA\, was selected to solve
for the \ion{O}{1} triple line abundance. Fig.~\ref{otrip} compares the
observed (histogram line) and SME synthesized (smooth curve) profiles of the
oxygen triplet lines in four stars, spanning a wide range of effective
temperatures and rotational velocities.  The deduced oxygen abundances are
presented in Table~\ref{triplet} [assuming a solar oxygen abundance O = 8.87 on
a logarithmic scale where H = 12.00 (Grevesse, Noels \& Sauval 1996)].  In
Table~\ref{triplet}, the stellar parameters $T_{\rm eff}$, $\log\,g$ and
micro-turbulent velocities are listed in Columns (2)--(4), respectively.
Columns (5)--(6) give, respectively, the ${\rm H}\alpha$ emission fluxes taken
from Mart\'{i}n \& Montes (1997) and the logarithmic X-ray luminosities divided
by the bolometric luminosity $L_{\rm bol}$, log $(L_{\rm X}/L_{\rm bol})$,
taken from Giampapa, Prosser \& Fleming (1998).  Column (7) gives the amplitude
of modulation in the V-band from Allain et al. (1996).  Columns (8)--(9) give,
respectively, the oxygen triplet line abundances and their uncertainties for
three stars, selected to represent different temperature regimes spanned by our
sample stars. Column (10) gives the oxygen triplet abundances deduced from the
equivalent width measurements.  A detailed description of the procedures used
to determine the stellar parameters and element abundances has been presented
in our Paper~I.
Full results, including stellar parameters and abundances of other individual
elements are presented there along with a detailed error analysis.  

To investigate the discrepancies between the oxygen abundances derived
from the permitted triplet lines on the one hand and those deduced from the
forbidden lines on the other, the simplest way is to directly compare the two
set of abundances.  However, our current set of spectra of IC\,4665 were
hampered by limited wavelength coverage (the stronger component $\lambda$6300
of the [O~{\sc i}] $\lambda\lambda$6300,6363 fell on the edge of the spectra)
and by relatively low S/N ratios, and we were unable to derive oxygen
abundances from the weak $\lambda\lambda$6300,6363 forbidden lines. Instead, we
have used the [\ion{O}{1}] $\lambda$6300 abundances deduced for dwarfs of the
Pleiades. This is reasonable, given that the two clusters have very similar
metallicities and ages. The Pleiades spectra of Wilden et al. (2002) did not
cover the oxygen triplet lines, yet the much higher S/N ratios of those spectra
allowed us to determine oxygen abundances using the [\ion{O}{1}] $\lambda$6300
forbidden line for six of them, i.e. H\,II\,129, 250, 522, 1039, 1298, and
2462. In the current analysis, we have adopted the stellar parameters determined by
Wilden et al.  (2002).  

The [\ion{O}{1}] $\lambda$6300 line is known to be contaminated by a weak
\ion{Ni}{1} line (Lambert 1978; Johansson et al. 2003).  To be consistent with
the previous work of Schuler et al. (2004), we modeled the blending
\ion{Ni}{1} line assuming a ${\rm log}\,gf = -2.31$ (Allende Prieto et al.
2001). Apart from the \ion{Ni}{1} line, between the narrow wavelength range of
6300.1--6300.5\,{\AA} where the [O~{\sc i}] $\lambda$6300 line falls, Kurucz
lists dozens CN lines on his web
site\footnote{http://kurucz.harvard.edu/LINELISTS/LINESMOL/}. However these
weak CN lines are believed to be only important for the continuum opacities and
so far no cases of significant contribution of CN lines to the observed
[\ion{O}{1}] $\lambda$6300 line strengths have been reported for solar type
stars (see, e.g., Allende Prieto et al. 2001; Schuler et al.  2004). We have
also considered possible contamination of the [\ion{O}{1}] $\lambda$6300 line
by telluric absorption.  From the line list of the solar spectrum of Moore et al.
(1966), the nearest telluric absorption line to the [\ion{O}{1}] line is at
6299.23\,\AA, in addition to two unidentified absorption lines at
6299.41 and 6299.59\,\AA. Pleiades has a heliocentric radial velocity of
$5.41\pm 0.37$\,km\,s$^{-1}$ (Kharchenko et al., 2005), leading to a blue-shift
of $-16.6$\,km\,s$^{-1}$, or $-0.35$\,{\AA} in wavelength at 6300\,{\AA} at the
time the spectra were taken, in good agreement with the average value of
0.3\,\AA\ determined from nearby strong lines in our spectra. We thus conclude
that it is unlikely that our measurements of the [\ion{O}{1}]
$\lambda$6300.30 line were affected by the nearby telluric absorption line and the two nearby lines of unknown
origin.

\begin{figure}
\plotone{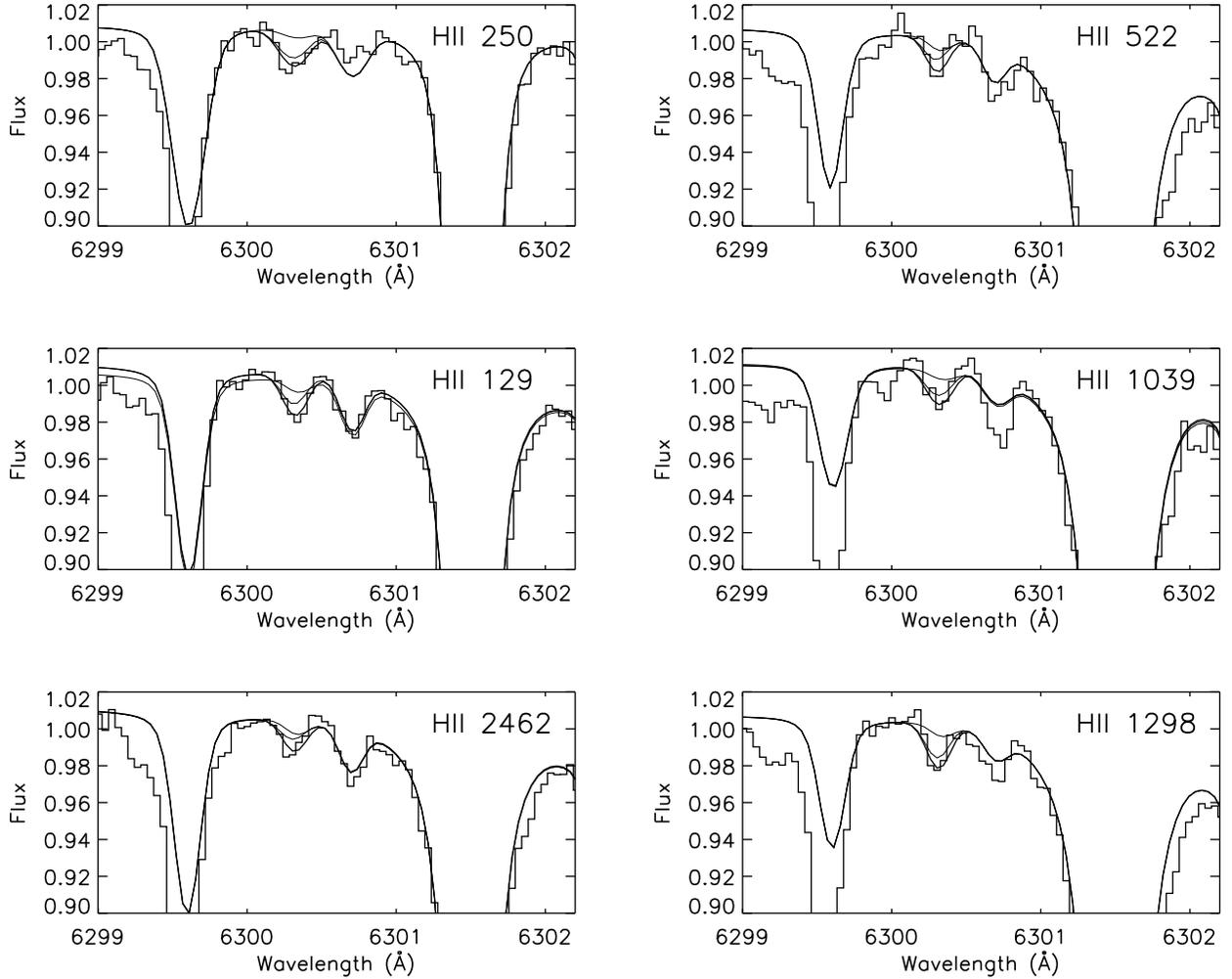}
\caption{Spectra of six Pleiades stars centered on the \ion{O}{1} $\lambda$6300
forbidden line. In each panel, also overplotted are three synthetic spectra
(smooth curves) showing, from top to bottom, the contributions from the
\ion{Ni}{1} $\lambda$6300.34 line, from the [\ion{O}{1}] $\lambda$6300.30
forbidden line and their sum, respectively.}
\label{pspectrum}
\end{figure}

The $\lambda$6300 feature was well fitted using the atomic data for the
[\ion{O}{1}] $\lambda$6300 line retrieved from the VALD (Vienna Atomic Line
Database; Piskunov et al. 1995; Ryabchikova et al. 1999; Kupka et al.  1999),
after taking into account contribution from the blending \ion{Ni}{1} line. As
such, we have not solved the atomic data for the [\ion{O}{1}] $\lambda$6300
line, but instead adopted those retrieved from the VALD. In order to reduce the
systematic errors, we re-measured the Ni abundances in the six stars using
\ion{Ni}{1} lines in the wavelength segments 5575--5595, 5636--5657, and
6107--6132\,\AA.  Atomic data for Ni lines in those wavelength segments were
solved using the National Solar Observatory solar spectrum (Kurucz et al.
1984). The Ni abundances ${\rm log}\,N({\rm Ni})$ derived by us (H\,II\,129:
6.11; H\,II\,250: 6.11; H\,II\,522: 6.15; H\,II\,1039: 6.13; H\,II\,1298: 6.15;
H\,II\,2462: 6.12) agree with those obtained by Wilden el al. (2002) within
0.04\,dex.  Contributions from the [\ion{O}{1}] forbidden line and from the
\ion{Ni}{1} line to the observed 6300\,\AA\, feature were synthesized
separately. The resultant synthetic spectra as well as their sum are plotted in
Fig.~\ref{pspectrum} and compared to the observed ones.  The oxygen abundances
thus derived from the $\lambda$6300 forbidden line are listed in
Table\,\ref{forbid}. 

Stellar elemental abundances are traditionally derived using the
equivalent width (EW) method. For comparison, we have also determined the
oxygen triplet abundances from the measured EWs of the \ion{O}{1} triplet
lines. The results are listed in Table~\ref{triplet}. The EWs were measured using multiple Gaussian fits. The oxygen
abundances were then derived using ABONTEST8, a program developed by P. Magain.
ABONTEST8 uses LTE plane-parallel model atmospheres given by Kurucz (1993). We
have adopted stellar parameters deduced from SME. Uncertainties of the results
were dominated by those in the measured EWs, which were estimated using the
equation given by Norris et al.  (2001). For cool stars in our sample, the
spectra had typical S/N's between 30--50 and uncertainties in the derived
oxygen abundances were estimated to be about 0.15--0.25\,dex. For warm stars,
the spectra had better S/N's and the uncertainties are smaller. Compared to SME
values, oxygen abundances derived from the EW method are found to be 0.15\,dex
lower on average, with a standard deviation of 0.11\,dex and a maximum
difference of 0.4\,dex. Larger deviations are obtained for stars with lower
temperatures. 

We have also measured the EWs of the [\ion{O}{1}] $\lambda$6300 line and
calculated oxygen abundances. Contributions from the blending \ion{Ni}{1} line
to the observed EWs of the [\ion{O}{1}] $\lambda$6300 line were estimated and
corrected for using the EWs of the \ion{Ni}{1} $\lambda$6300.34 line measured
on SME model spectra. The results are listed in Table~\ref{forbid}.  Within the
uncertainties, they are consistent with the SME values.  The mean value of the
EW-based abundances is $-$0.05\,dex, only 0.04\,dex lower than the average of
the SME-based abundances.  Note that the $\lambda$6300 absorption line falls in
the extended wings of two nearby strong \ion{Si}{1} $\lambda$6299.60 and
\ion{Fe}{1} $\lambda$6301.50 lines. The presence of these strong lines makes the
placement of local continuum rather difficult, leading to large uncertainties
in EW measurements, in particular for weak lines such as the [\ion{O}{1}]
$\lambda$6300 line. By contrast, these strong lines are well modeled 
in SME and therefore abundances deduced using the SME method are less 
affected by their presence.

\section{Discussion}

\begin{figure}
\plotone{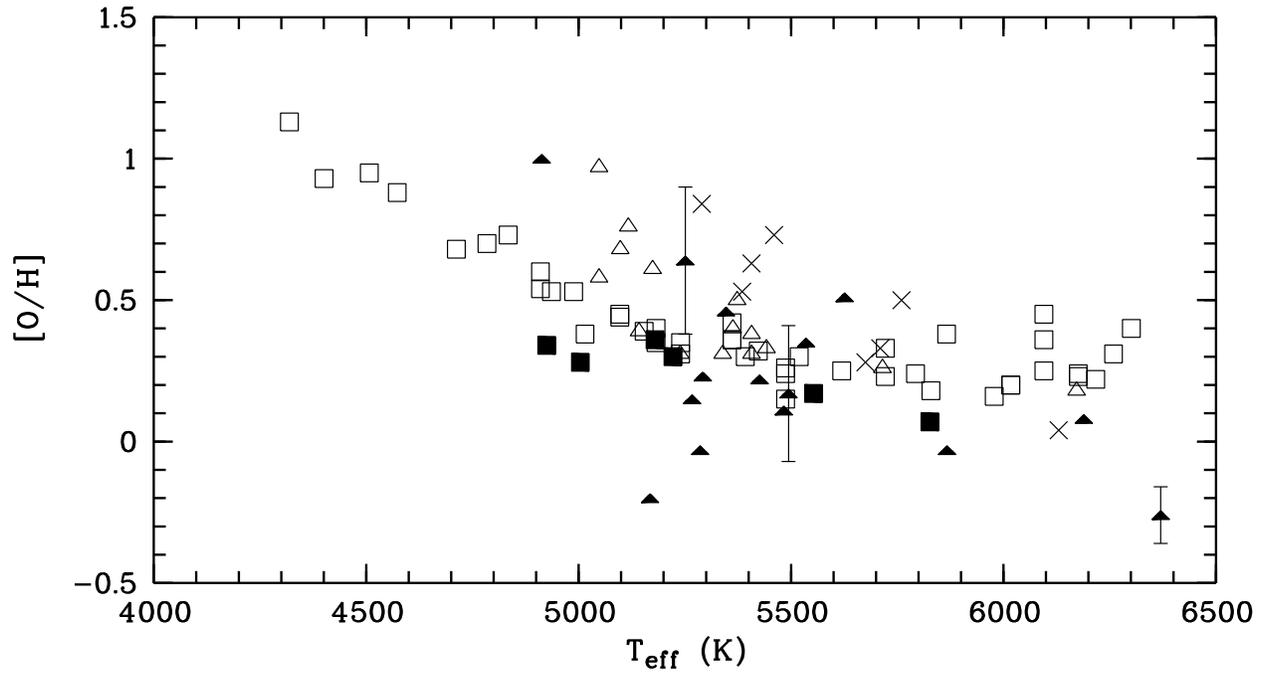}
\caption{Oxygen triplet line abundances of IC\,4665 dwarfs (filled triangles)
plotted against effective temperature. Typical errors are shown for three
selected stars of representative effective temperatures. Also plotted are
oxygen triplet line abundances of Pleiades (open triangles) and M\,34
dwarfs (crosses) taken from Schuler et al. (2004), Hyades dwarfs (open 
squares) from Schuler et al. (2006a) and UMa dwarfs (filled squares) from King
\& Schuler (2005)}
\label{ic4665}
\end{figure}

Oxygen abundances relative to the solar value for a sample of 15 IC\,4665
stars, derived from the triplet permitted lines, are plotted in
Fig.~\ref{ic4665} against effective temperature $T_{\rm eff}$. Typical error
bars for three selected stars of representative values of $T_{\rm eff}$ are
indicated in the diagram. The upper boundary of the abundance show a clear
trend with $T_{\rm eff}$, increasing dramatically by about 1.3\,dex as $T_{\rm
eff}$ decreases from $\sim 6400$ to 4900~K.  For stars of intermediate $T_{\rm
eff}$ ($\sim 5400$~K), the abundances exhibit a large scatter.  A similar trend
is also found for EW-based triplet line oxygen abundances, which show
variations over a range of 1.0\,dex, only slightly smaller than the spread of
1.3\,dex for the SME-based values. This demonstrates that the observed scatter
is not an artifact caused by our SME analysis method. The upper boundary trend
is similar to what found previously for dwarfs in Pleiades and M\,34 by Schuler
et al. (2004), and to a less extent for the cool dwarfs ($\la$6000\,K) in
Hyades by Schuler et al. (2006a) and UMa by King \& Schuler (2005) based on the
method of EW analysis.  However, no large spread within small $T_{\rm eff}$
intervals has been reported in the other open clusters as in IC\,4665.  

\begin{figure} 
\plotone{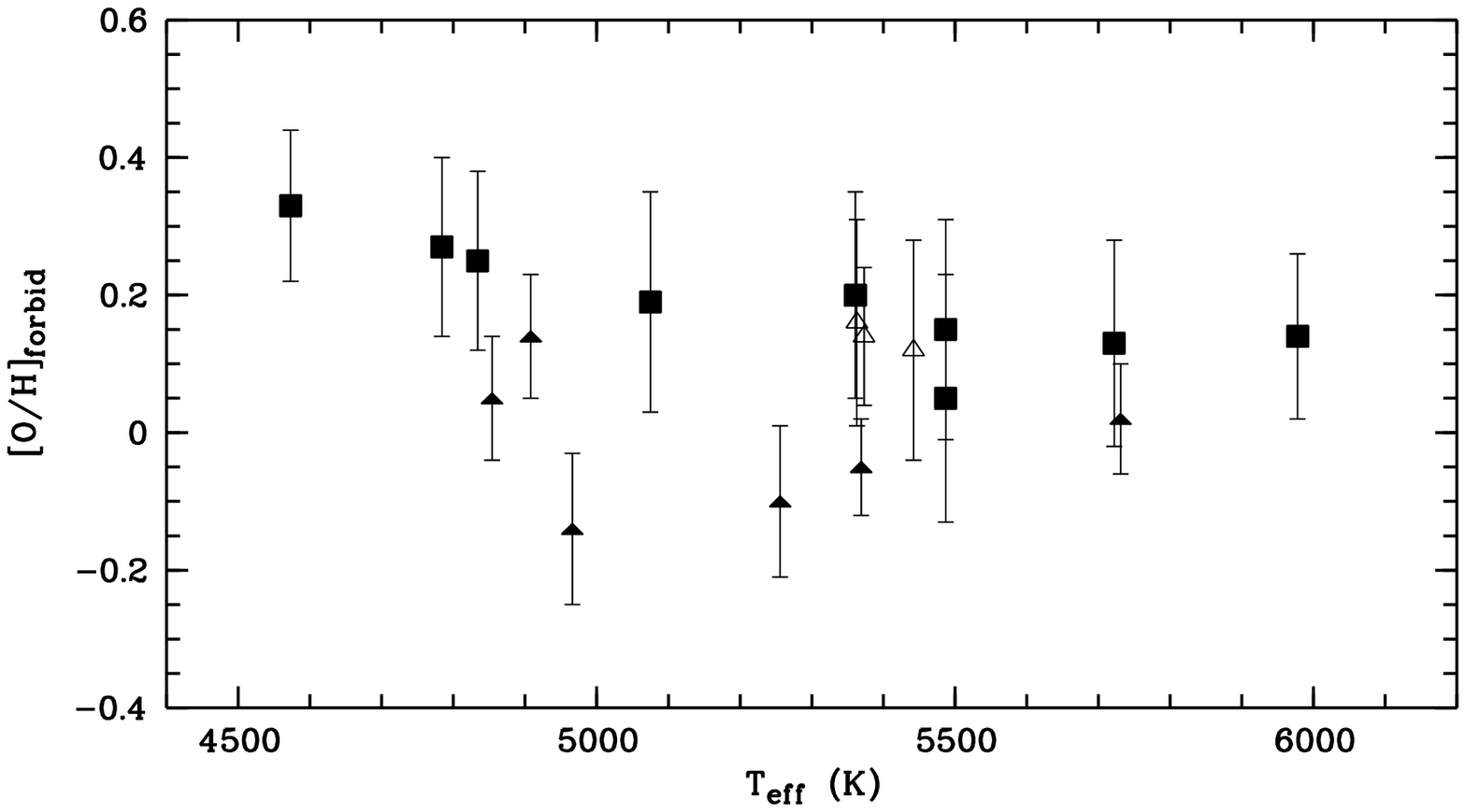} 

\caption{[\ion{O}{1}] forbidden line abundances of Pleiades dwarfs plotted
against effective temperature. Filled triangles are our results and open
triangles are values taken from Schuler et al. (2004). Also plotted are
[\ion{O}{1}] forbidden line abundances of Hyades (filled squares) taken from
Schuler et al. (2006b).}

\label{pleiades} 
\end{figure}

In contrast to the triplet abundances, no such trend is found for oxygen
abundances deduced from the $\lambda$6300 forbidden line. In
Fig.~\ref{pleiades} we plot the [\ion{O}{1}] abundances of six Pleiades stars
in our sample against $T_{\rm eff}$ and another three Pleiades stars previously
analyzed by Schuler et al. (2004). Also included in the figure are oxygen
forbidden line abundances of 8 Hyades stars taken from Schuler et al. (2006b)
based on stellar atmosphere models with overshoot.  Fig.~\ref{pleiades} shows
clearly that oxygen abundances of Pleiades stars derived from the [\ion{O}{1}]
forbidden line have small scatter over the entire range of $T_{\rm eff}$, from
$\sim 4800$ to 5800~K.  The three stars of Schuler et al. (2004) yield an
average [\ion{O}{1}] abundance of 0.14\,dex, which is 0.15\,dex higher than our
mean abundance of $-$0.01\,dex.  The measured EWs of the [\ion{O}{1}]
$\lambda$6300 line in the sample of Schuler et al. (2004) are generally
1--2\,m\AA\ stronger than ours.  However, given the estimated observational
error of $\sim$3\,m\AA\ from Schuler et al. (2004), the EW differences seem not
so significant. I prefer that the abundance difference between the two works
are due to the systematic errors.  Although the three coolest dwarfs
(4573$\leq$ $T_{\rm eff}$ $\leq$ 4834\,K) seem to increase with decreasing
$T_{\rm eff}$, [\ion{O}{1}] abundances of stars in Hyades do not show spreads
within the stated uncertainties.  The Hyades [\ion{O}{1}] abundances are still
almost constant with a mean value of 0.19\,dex which is consistent with the
metallicity of Hyades ([Fe/H]=0.13; Paulson et al.  2003).  

Under the assumption that member stars of a given open cluster all have the
same chemical composition, an assumption that is indeed supported by
measurements of oxygen abundances using the [\ion{O}{1}] forbidden line, it is
probably reasonable to assume that the dramatic increase of the oxygen triplet
line abundances with decreasing $T_{\rm eff}$ as shown in Fig.~\ref{ic4665} is
an artifact and therefore the triplet line abundances do not represent the true
stellar surface oxygen abundances. Rather, it seems that they must have been
affected by some systematic effects which are a strong function of\, $T_{\rm
eff}$.  Possible mechanisms that might be responsible for the abnormal triplet
line oxygen abundances include systematic uncertainties in the abundance
determinations, NLTE effects, granulation, surface activities (e.g.,
chromospheric activities, stellar spots). We examine these possibilities one by
one in the following subsections. 

\subsection{Abundance determination uncertainties}

In Paper~I, we discuss our method of effective temperature determination and
compare the results with those deduced from the photometric method using color
indices. In our analysis based on SME, we use about one hundred iron lines of
different excitation energies when determining the effective temperature and
require that they yield self-consistent element abundances.  SME however
does not directly force a balance in the excitation potential or ionization
states. Instead, it uses a Levenberg-Marquardt algorithm to minimize $\chi^2$
in a synthetic spectral model of the observed spectrum. During this process,
each line gives its own independent verdict regarding the free parameters,
regardless whether SME is used to solve stellar parameters such as $T_{\rm
eff}$, $\log\,g$ or element abundances. The standard deviations of the ensemble
of hundreds of lines provides estimates of the uncertainties in the derived
parameter.  On the other hand, as previously found by Valenti \& Fischer
(2005), this algorithm generally grossly underestimates the true uncertainties.
The lack of a trend in [Fe/H] with $T_{\rm eff}$ (see Paper~I) gives a
post-check that the analysis has been properly carried out, in other words, we
have $T_{\rm eff}$ properly determined. The method has the advantage that the
results are insensitive to potential uncertainties in the stellar atmospheric
models and in the photometric measurements of the color indices. In the
spectroscopic method, the effective temperature and element abundances are
determined from the same set of observations, thus avoiding any potential
uncertainties introduced by combining observations obtained with different
instruments and telescopes in order to calibrate the color indices.
 
In Paper~I, we estimate that SME temperatures are accurate to $\pm 100$~K.
The estimate was based on the fact that the differences between temperatures
derived from SME and from $(B-V)$ colors yield a standard deviation of only
141~K and the fact that we believe that spectroscopic temperatures are more
accurate than color temperatures. By adding in quadrature the measurement error
in a single observation, and the numerical error yielded by the analysis
algorithm (the standard deviation of final parameter values about the mean for
each observation, due only to differences in initial values), Valenti \&
Fischer (2005) estimated the uncertainties for temperatures obtained by fitting
a single observation of a generic star to be 44~K. Their work was based on an
improved version of SME. Given that our spectra generally have lower S/N's than
theirs, it is probably realistic to assume that our estimates of $T_{\rm eff}$
are accurate to approximately 100~K. Values of $\log\,g$ were calculated using
Eq.\,(16.2) of Gray (1992) and assigned a nominal uncertainty of 0.1\,dex. Our
estimates of typical uncertainties of 100~K for $T_{\rm eff}$, 0.1\,dex for
$\log\,g$ and 30\% for $v_{\rm mic}$ are in line with what assumed by Wilden et
al. (2002) who also adopt SME in their analysis.

The effects of uncertainties in individual stellar parameters on the derived
element abundances have been analyzed in Paper~I. The oxygen triplet permitted
lines are sensitive to effective temperature, surface gravity, and
microturbulent velocity determinations. In order to obtain an estimate of the
relative error of the triplet line oxygen abundance, we recalculated the
abundance by varying the effective temperature by amounts of $\pm 100$\,K,
${\rm log}\,g$ by $\pm 0.1$\,dex and the microturbulent velocity by $\pm 30$\%.
Calculations were carried for three stars (here we changed P\,332 in Paper~I to
P\,71 as the former has been excluded from the sample. The calculated
uncertainty difference between P\,332 and P\,71 would not affect our results in
Paper~I.), selected to represent the full range of effective temperature of
stars in the current sample.  Uncertainties introduced in the process of line
profile fitting were also estimated and incorporated into the final error
budget by varying the elemental abundance and then analyzing its effects on the
residuals of the fit.  The results, tabulated in Table~\ref{triplet}, show that
even though the total formal errors of oxygen abundances derived from the
triplet lines can reach as much as 0.26\,dex for stars of effective
temperatures lower than 5500\,K, they are insufficient to account for the
observed systematic variations of the oxygen triplet line abundances of more
than 1\,dex as shown in Fig.~\ref{ic4665}.

Valenti \& Fischer (2005) have used SME Version 2 to derive stellar parameters
and elemental abundances for a sample of over 1000 Galactic field stars, mainly
of dwarfs of spectral types F, G and K, using high quality spectra obtained as
a by-product of the California \& Carnegie Planet Search program. A thorough
error analysis was also carried out and some trends of the deduced elemental
abundances as a function of $T_{\rm eff}$ were found.  They interpreted such
trends as spurious and used polynomial fits to remove them.  In all cases, the
corrections amount to less than 0.2\,dex for the temperature range from 4900 to
6400~K. Such trends are not unique to SME but also exist in other methods of
abundance analysis, such as those based on EWs (e.g.  Schuler et al. 2003).
But more importantly, the range of variation of the triplet line oxygen
abundances observed in Fig.~\ref{ic4665} is so large such that we deem it
highly unlikely that the variations are of similar origins of the spurious
trends found in Valenti \& Fischer (2005).

Uncertainties in the derived [O/H]$_{\rm forbid}$ abundances are dominated by
photon shooting errors as a result of the limited S/N ratios of the weak
diagnostic line. For oxygen abundances derived from the forbidden line,
uncertainties caused by errors in $T_{\rm eff}$ and in ${\rm log}\,g$ are
essentially identical for all the sample stars and amount to 0.02 and 0.04\,dex,
respectively, for stars in our sample. The [\ion{O}{1}] $\lambda6300$ line is very 
weak and therefore not affected by microturbulent velocity determinations. 
The total errors in the deduced
abundances are obtained by adding photon shooting errors and those introduced
by uncertainties in $T_{\rm eff}$ and in ${\rm log}\,g$ in quadrature.  The
results are listed in the last column of Table\,\ref{forbid}.
Fig.~\ref{pleiades} shows that the abundances deduced from the [\ion{O}{1}]
forbidden line have a much smaller scatter compared to those derived from the
\ion{O}{1} triplet lines. The values vary over a range of 0.28\,dex with a
standard deviation of 0.1\,dex, comparable to the uncertainties of
0.07--0.11\,dex for individual measurements. The sample yields an average [O I]
abundance of $-$0.01\,dex, i.e. nearly identical to the solar value.

\subsection{NLTE effects}

The \ion{O}{1} triplet lines are known to suffer from the NLTE effects. For a
wide range of stellar atmosphere parameters, the NLTE effects will lead to
overestimated oxygen abundances (e.g. Gratton et al. 1999; Takeda 2003). The
effects are however small, less than 0.1\,dex for cool ($T_{\rm eff} < 6000$~K)
dwarfs of approximately solar metallicity. For stars of higher temperatures,
the NLTE effect are expected to be larger (Takeda 2003).  Thus the trend caused
by the NLTE effects is exactly opposite to what observed.  It seems highly
unlikely that the nominal NLTE effects are responsible for the trend observed
in Fig.~\ref{ic4665}. 

Observations (e.g., King \& Boesgaard 1995) show that for stars with $T_{\rm
eff} \ga 6200$~K triplet-based oxygen abundances are enhanced compared to
forbidden line-based values, in agreement with the theoretical predictions of
Takeda (2003) for main-sequence stars of $T_{\rm eff} \sim 6500$~K. On the
other hand, the enhancement is small (about 0.18\,dex) and is likely to be lost
in the huge variations depicted in Fig.~\ref{ic4665} even if it exists in
IC\,4665. We note that P\,19 ($T_{\rm eff} = 6370$~K), the warmest star in our
sample, has the lowest oxygen abundance. Its spectrum is of high quality, with
a S/N of 150.  SME and EW analyses give very similar abundances, [O/H] $=
-0.31$ and $-0.26$\,dex, respectively. According to the theoretical
calculations of Takeda (2003), the EWs of the oxygen triplet lines measured for
P\,19 implies an NLTE correction of $\sim -0.2$\,dex.  Applying this correction
will lead to very low oxygen abundance of $\sim$ $-$0.5\,dex for P\,19. Even
the existence of a binary companion is not likely to affect its oxygen
abundance.  It appears to us that P\,19 does seem to have an abnormally low
oxygen abundance compared to other member stars of the cluster.

In their analysis of the open cluster M\,34, Schuler et al. (2003) find that Si
abundances derived from lines of high excitation potentials ($\chi \sim 6
$\,eV) increase with decreasing temperature, whereas abundances of Fe, Ti, Cr,
Ca, Al and Mg, all derived from lines of relatively low excitation potentials
($\chi \approx 2$--4\,eV), decrease. They suggest that
overexcitation/ionization, caused by deviations from the LTE in cool stars, may
be responsible for the observed abundance trends.  Similar trends are also
observed in our results of abundance analysis for IC\,4665 (Paper~I) and those
for the Pleiades by Wilden et al. (2002), where it is found that as $T_{\rm
eff}$ decreases, Si abundance, deduced from lines with $\chi$ $\sim$ 6\,eV,
increases, whereas Cr abundance, derived from lines of $\chi$ $\sim$ 
0--4\,eV, decreases.  There are however exceptions. For example, in IC\,4665,
Ti abundance determined from lines of $\chi \sim 0$--2\,eV increases as $T_{\rm
eff}$ decreases. Given the relatively high excitation potential of the oxygen
triplet lines, it is difficult to rule out the possibility that the observed
strong increase of the oxygen triplet line abundances with decreasing $T_{\rm
eff}$ is caused by overexcitation/ionization as proposed by Schuler et al.
(2003).  Further quantitative investigations of this scenario, i.e. deviations
from the LTE can indeed result in overexcitation/ionization in cool stars, and
if so, its effects on elemental abundances derived from lines of different
excitation potentials, are highly desirable. 

\subsection{Surface activities}

\subsubsection{Chromospheric activities}

Takeda (1995) suggests that the oxygen triplet line formation could be affected
by chromospheric activities. In their analysis of 14 single-lined RS CVn
binaries, Morel \& Micela (2004) find suspiciously high \ion{O}{1} triplet line
abundances, by as much as 1.8\,dex, in chromospherically active systems.  They
plot oxygen abundances deduced from the triplet lines and from the forbidden
line against chromospheric activity indicators, including strength of the
\ion{Ca}{2} H$+$K lines and X-ray flux. They find that for both RS CVn binaries
and stars of young open clusters, as the level of chromospheric activity
increases, the \ion{O}{1} triplet line abundance increases, whereas those
derived from the forbidden lines remains nearly constant around the solar
value.   The results of Morel \& Micela strongly suggest that the steep
rise of the \ion{O}{1} triplet line abundance with decreasing $T_{\rm eff}$ is
probably caused by chromospheric activities.
However, one difficulty of the study was that all stars in their RS CVn binary
sample have effective temperatures near or below 5000\,K. As a result, Morel \&
Micela were unable to rule out the possibility that the observed large
variations of the oxygen triplet line abundances with $T_{\rm eff}$ was not
caused by inadequacy in the Kurucz LTE atmospheric models for late K-type stars.
On the other hand, a recent analysis of six K dwarfs and subgiants by 
Affer et al. (2005) shows that the usage of LTE Kurucz model 
atmospheres does not lead to large variations of oxygen triplet line 
abundances in those old K stars which have low surface activities.

From an analysis of more than 200 active stars in open clusters and in the
field, Messina et al. (2001) concluded that the strength of activities
increases from zero-age, reaching maxima around the age of Pleiades ($\sim$70
Myr), and then declines. If surface activities account for the observed
behavior of the triplet line abundance, then one would expect a correlation
between the linear fitting slope of the oxygen triplet line abundances as a
function of effective temperature and the age of the open clusters.  We
calculated the slopes for stars with $T_{\rm eff}$ below 6000~K in each
cluster.  Stars with  $T_{\rm eff} >$ 6000~K are excluded because they are
believed to be affected by NLTE effect (e.g., Schuler et al. 2006a).  From the
currently rather restricted amount of data, we find almost identical absolute
values of the slopes for the two younger open clusters Pleiades ($\sim$70\,Myr)
and M\,34 ($\sim$250\,Myr) of ($-8.2\pm2.1$) and ($-8.8\pm3.0)\times10^{-4}$
respectively, compared to the much smaller slopes of ($-3.0\pm0.7$) and
($-5.1\pm0.4)\times10^{-4}$ for the older clusters UMa ($\sim$600\,Myr) and
Hyades ($\sim$600\,Myr).  The slope of IC\,4665 ($\sim$35\,Myr) is also much
small ($(-4.4\pm2.0)\times10^{-4}$), comparable to those of UMa and Hyades.
The result seems to be consistent with the prediction of Messina et al. (2001).
We should notice that the small slope of IC\,4665 is due to the large scatter
around 5400\,K. Comparing the triplet abundances variation of different
clusters within the same $T_{\rm eff}$ range in Fig.~\ref{ic4665}, IC\,4665
stars show a much higher oxygen abundance upper boundary than the older
clusters UMa and Hyades.  Although inconclusive, these results seem to suggest
that age-related effects (e.g., surface activities including chromospheric
activity, stellar spots and flares) is influencing triplet abundance
determinations for young open clusters. 

In order to clarify the role of chromospheric activity in the behavior of
oxygen triplet abundance in young open clusters, we calculated a factor R based on the emission of chromospheric activity sensitive lines ${\rm H}\alpha$ and
\ion {Ca}{2} IRT $\lambda\lambda$8498,8662.  Montes \& Mart\'{i}n (1998)
provide a set of standard quiet stars observed at high-resolution thus we can
use them in the application of the spectral subtraction technique to obtain the
active-chromosphere contribution to these activity-sensitive lines.  
We calculated R from the fluxes (F) of ${\rm H}\alpha$ and
\ion {Ca}{2} IRT $\lambda\lambda$8498,8662 by  
\[
R = {F_{\rm IC\,4665}-F_{\rm standard} \over F_{\rm standard}}
\] 
The standard quiet stars are chosen from the sample of  Montes \& Mart\'{i}n (1998) to have similar $(B-V)$ as the IC\,4665 stars.
The results are listed in Table~\ref{R}. Observational errors are also estimated and listed in columns (5)-(7).
Column (8) of Table~\ref{R} gives the errors of the \ion{O}{1} triplet abundance for each star, calculated from 
a linear fitting equation of $T_{\rm eff}$ and the uncertainties of the three 
representative stars.

\begin{figure}
\plotone{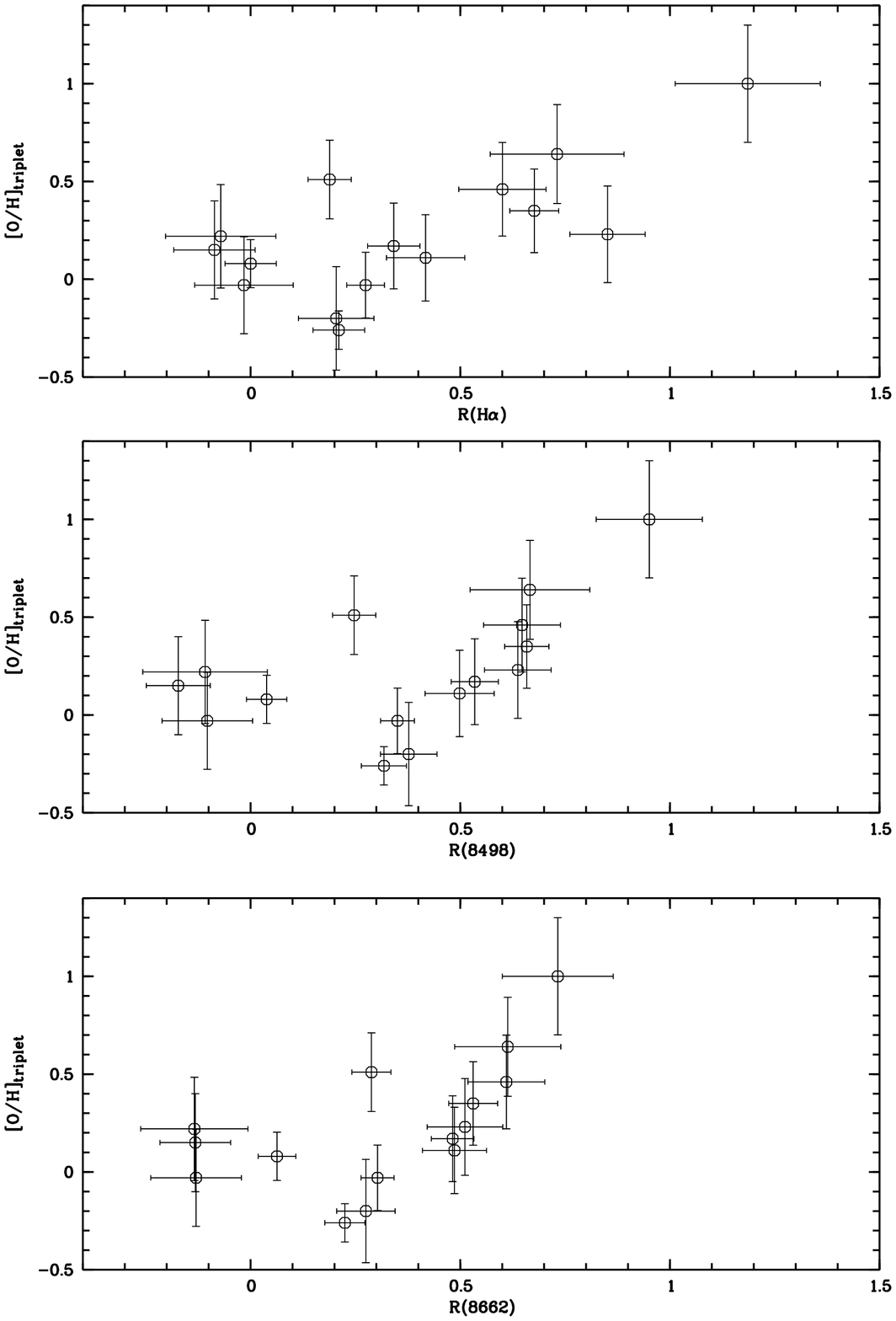}
\caption{Oxygen triplet line abundance as a function of H$\alpha$ and \ion{Ca}{2} IRT line $\lambda\lambda$8498,8662 emission factor R for IC\,4665 sample stars.}
\label{Hairt}
\end{figure}

The oxygen triplet abundance is plotted against the three activity indicators
R(H$\alpha$), R(8498), and R(8662) in Fig.~\ref{Hairt}. The linear correlation
coefficients of the triplet line abundances with the activity indicators are
calculated and listed in Table~\ref{active}. Given the correlation coefficients
of 0.69, 0.55 and 0.57 of the oxygen triplet abundances with the three
indicators respectively, only marginal correlations could be suggested for
them. However, from Fig.~\ref{Hairt}, we find that for stars with high R (high
activity), there does exist a clear correlation between triplet abundances and
the activity indicators, whereas the stars with low R show totally no
correlation.  
Given that the stars which show near-zero or even nagative R meaning that they
should have stable chromospheric activity also have oxygen abundances
which are consistent with the [\ion{O}{1}] abundances within uncertainties,
it is likely that the oxygen triplet abundance variation of the four stars with
R smaller than 0.1 is just due to errors other than activities.  If we
excluded these 4 stars whose triplet abundances are not affected by
chromospheric activities, the linear correlation coefficients of oxygen triplet
abundances with the activity indicators would increase to 0.76, 0.73 and 0.81
for the remaining stars, which are significant at the confidence levels of
99.99\%, 99.98\% and 99.99\%, respectively. These are much convinced values to
support that chromospheric activity is responsible for the oxygen triplet
abundance behaviors. The only point which prevents the correlation coefficients
to increase over 0.90 is P\,107, a fast rotator whose emission level may be
underestimated.

Besides our work, X-ray luminosities, normalized to unit bolometric luminosity,
are available for three of our stars, P\,60, P\,71 and P\,100 (Giampapa,
Prosser \& Fleming 1998).  ${\rm H}\alpha$ emission fluxes, derived using the
technique of spectral subtraction, are available for seven of them (Mart\'{i}n
\& Montes 1997). The two chromospheric activity indicators are listed in
Table~\ref{triplet}. 
Although the X-ray data are scarce, they do point to a positive correlation
with oxygen abundances determined from the triplet lines.  The linear
correlation coefficient is 0.98 with a statistic significance of 87\%.  Whereas
the ${\rm H}\alpha$ emission data show a large scatter without a clear
correlation between the oxygen triplet abundances and the ${\rm H}\alpha$
emission fluxes (the linear correlation coefficient is 0.34). Among the stars,
P\,94 has a low oxygen triplet line abundance and a low H$\alpha$ emission
flux. If we exclude the star with the lowest ${\rm H}\alpha$ emission flux
(P\,107), then the remaining six stars do show a positive correlation with a
linear correlation coefficient of 0.78 with a significance of 93\%. 

In addition to IC\,4665, we have also calculated the linear correlation
coefficients of oxygen triplet line abundances with different surface activity
indicators (H$\alpha$, X-ray, \ion{Ca}{2} H+K lines and infrared triplet lines,
and the amplitude of modulation in the V-band) for other open clusters.  The
results are listed in Table.~\ref{active}. In the table, the last column gives
the references for the data used. Values in parentheses are numbers of stars
used in the calculations.  For Pleiades stars, the oxygen triplet abundances
are found to be mildly correlated with the two chromospheric activity
indicators, X-ray luminosity and the \ion{Ca}{2} infrared triplet (IRT) line
strength, but not with H$\alpha$ emission.  For M\,34 stars, the triplet
abundances are mildly correlated with the \ion{Ca}{2} IRT line strength, but
not with H$\alpha$ emission.  For UMa stars, the oxygen triplet abundances are
mildly correlated with the \ion{Ca}{2} H+K line strength. For Hyades stars, no
correlations are found between oxygen triplet abundances and the X-ray emission
or the \ion{Ca}{2} H+K line strength.  Except for IC\,4665 where the oxygen
triplet abundances are found to be correlated with the X-ray luminosities in
three stars, no obvious correlations are found for other open clusters.  The
correlations between oxygen abundances in different clusters and different
chromospheric activity indicators are thus too ambiguous to draw a firm
conclusion.  One should remember that in all cases but our work, the stellar
spectra and the chromospheric activity indicators were not measured at the same
epoch. Therefore, we really do not know the activity levels of the stars at the
time the spectra were taken.  Simultaneous observations of oxygen triplet lines
and chromospheric activity indicators are essential to clarify the situation. 

In the previous works, one fact that made the situation further complicated is
that both oxygen triplet abundances and chromospheric activity indicators are
correlated with $T_{\rm eff}$.  Thus the triplet abundance-activity correlation
may be a consequence of the $T_{\rm eff}$ correlations, prohibiting a strong
conclusion from being made about a causal relationship between activity
indicators and anomalous triplet abundances.  However, in our sample of
IC\,4665, the correlation between oxygen triplet abundance and $T_{\rm eff}$
are not so monotonous (the linear correlation coefficient is 0.53) as there is
a large spread around 5400\,K. At the same time, the correlation of the
abundance and chromospheric activity indicators are much better for stars which
show chromospheric activity emissions (the coefficient $\sim$ 0.75).  This
would be a strong evidence to solve the degeneracy between the triplet
abundance/activity correlation and the abundance/$T_{\rm eff}$ correlation.
The results support chromospheric activity as the main reason for the abnormal
oxygen triplet abundance observed in young open clusters. Furthermore,
theoretical models investigating the possible effects of stellar activities on
line formation and element abundance determinations are still needed to totally
disentangle the problem.

\subsubsection{Stellar spots}
 
Atmospheric perturbations (e.g., stellar spots, flares) have been invoked to
explain at least partially the large Li abundance spread in cool members of
young open clusters (Barrado y Navascu\'{e}s et al. 2001; Ford et al. 2002;
Xiong \& Deng 2005). Surface activities such as variations in spot filling
factor may affect element abundance determinations in two ways. Firstly, they
can cause color anomalies (Stauffer et al. 2003) and thus introduce
uncertainties in effective temperature determinations.  In SME, effective
temperatures are determined directly from the absorption line spectrum, rather
than from the stellar energy distribution, i.e. color indices. Thus to the
first approximation, color anomalies caused by atmospheric perturbations are
irrelevant to our analyses based on the technique of SME.  Alternatively,
atmospheric perturbations can influence directly the line formation process.
Using a simple model consisting of arbitrarily chosen line flux contributions
from cool and hot spots, Schuler et al. (2006a) found that stars with spots of
different temperatures is capable to reproduce the observed equivalent widths
for 3 Hyades stars with different $T_{\rm eff}$ and triplet abundances.
However, how such effects can lead to a correlation between oxygen triplet
abundances and $T_{\rm eff}$, as depicted in Fig.~\ref{ic4665}, is still not
clear.  

Based on the model of Bouvier et al. (1993), the observed amplitude of
modulation in the V-band could be a crude estimation of the area of the spot
coverage on the surface.  The amplitudes of the V-band variation for 5 of our
sample stars given by Allain et al. (1996) are listed in Table ~\ref{triplet}.
The linear correlation coefficient between them and the triplet abundances is
0.46 with a statistic significance of 56\%, which leaves the correlation
between spots and the triplet abundances still unrecognized. However, we should
notice that again the photometric and spectroscopic observations are not
carried out at the same time.  Simultaneous observations are still needed to
draw a clear conclusion.    

Various studies have indicate that there are correlations between the
activities of different stellar atmosphere layers (e.g. Messina et al.
2003). Therefore even if we obtained the correlation between the
oxygen triplet abundances and chromospheric activity levels, we cannot
tell if the correlation are due to chromospheric activity or
activities of the other layers (such as spots).

\subsection{Granulation corrections}

Photospheric temperature fluctuation is another controversial factor that may
affect spectroscopic abundance determinations. Detailed calculations of
granulation abundance corrections have been carried out in recent years.  The
3D abundance corrections to be applied to the 1D solutions for individual
oxygen lines have been given by several works (e.g., Kiselman \& Nordlund
1995; Asplund 2001; Allende Prieto et al. 2001; Asplund \& Garc\'{i}a P\'{e}rez
2001; Nissen et al. 2002).  Kiselman \& Nordlund (1995) report negligible 3D
effects for the high-excitation \ion{O}{1} triplet lines, which form deep in
the photosphere.  Similar results are found by Nissen, Primas \& Asplund
(2001). Nissen et al. (2002) show that 3D correction for the [\ion{O}{1}]
forbidden line $\lambda$6300 is negligible in most cases, except in low
gravity/metallicity atmospheres where it can increase to about 0.2\,dex.
Steffen \& Holweger (2002) study the problem of LTE line formation in the
inhomogeneous solar photosphere based on detailed 2-dimensional radiation
hydrodynamics simulations of the convective surface layer of the Sun. By means
of a strictly differential 1D/2D comparison of the emergent equivalent widths,
they obtain the granulation abundance corrections for individual lines to be
applied to the standard abundances determined based on a homogeneous 1D model
atmosphere. For the oxygen triplet lines, the correction amounts to about
0.01\,dex. By constructing a realistic time-dependent, 3D, hydrodynamic model
of the solar atmosphere, Asplund et al.  (2004) determine the solar
photospheric oxygen abundance from a variety of diagnostic lines, including the
[\ion{O}{1}] forbidden lines, the \ion{O}{1} triplet permitted lines as well as
OH vibration-rotation lines and the OH pure rotation lines.  In the case of the
\ion{O}{1} triplet permitted lines, 3D NLTE calculations have been performed,
revealing significant departures from the LTE as a result of photon losses in
the lines. The NLTE effects due to this process yield corrections that amount
to 0.2--0.3\,dex. The differences between the 1D and 3D models are however
found to be small, less than 0.1\,dex. From these detailed calculations, it
seems that granulation corrections are probably unlikely to be able to account
for the large variations in the oxygen triplet line abundance discuss in the
current paper, at least in solar type stars.  Calculations of granulation
corrections for other types of star have not been reported and are preferred,
especially for late type stars.

\section{Summary}

We have found a dramatic increase in the oxygen triplet abundance upper
boundary with decreasing effective temperature in the cool dwarfs of young open
cluster IC\,4665, similar to what previously observed in Pleiades and M\,34 by
Schuler et al. (2004), in Hyades by Schuler et al. (2006a) and in UMa by King
\& Schuler (2005).  By contrast, oxygen abundances derived from the
[\ion{O}{1}] $\lambda$6300 forbidden line are found to be constant in Pleiades
and Hyades.  It seems that the [\ion{O}{1}] $\lambda$6300 forbidden line is
relatively free from the various processes that may have affected abundances
determined from the \ion{O}{1} triplet lines and is therefore a better
abundance indicator.

At the present moment, the uncertainties in the measured values of
[O/H] are too large to place any meaningful planet-formation
constraint based on the oxygen abundance of cluster member stars.
Under the assumption that oxygen abundance is homogeneous in a given
cluster, an assumption that is supported by abundance determinations
using the [\ion{O}{1}] forbidden line, we have investigated various
possible mechanisms that may be responsible for the observed trend and
scatter of oxygen triplet line abundances. Although the \ion{O}{1}
triplet lines are sensitive to stellar parameters, we show that the
possible uncertainties in our parameter determinations are unlikely to
be sufficient to explain the more than 1\,dex variations in oxygen
triplet line abundances.  The possible effects of canonical NLTE as
the dominant cause of the problem is ruled out as they predict a trend
of oxygen abundance as $T_{\rm eff}$ varies that is exactly opposite
to what observed.  Available calculations of granulation corrections
show that the effects are generally small, at least for solar type
stars, and therefore may contribute little to the observed spreads of
oxygen triplet line abundances.  The variation of oxygen triplet
abundances as a function of effective temperatures are found to be
larger in younger open clusters and smaller in older clusters.
Age-related effects stellar surface activities are then suggested to
blame for the large spreads of oxygen triplet abundances.  This
assumption are supported by the correlation analysis between the
triplet abundance and the simultaneous observation of H$\alpha$ and
\ion{Ca}{2} IRT emissions which, are indicators of stellar
chromospheric activity levels. The lack of a monotonous correlation
between $T_{\rm eff}$ and the triplet abundance implies that the the
triplet abundance/activity correlation is not a consequence of the
abundance/$T_{\rm eff}$ correlations.  We could not exclude the
possibility that stellar spots play a role in the abnormal behavior of
the detected oxygen triplet abundances.

\begin{acknowledgements} The authors wish to thank Dr. Debra Fischer for her
expert assistance in the usage of SME. We thank Dr. Frank Grupp for his valuable suggestions on this paper. ZXS and XWL acknowledge Chinese NSFC
Grant 10373015.  

\end{acknowledgements}

\begin{table} 
\tabcolsep 3pt
\centering
\caption{Stellar parameters and oxygen triplet line abundances of IC\,4665 dwarfs}
\label{triplet}
\begin{tabular}{lccccccrcr} 
\hline
\hline
\noalign{\smallskip}
Star & $T_{\rm eff}$(SME)& $\log\,g$ & $v_{\rm mic}$ & $F({\rm H\alpha})^{a}$ &
 log & A(V)$^{c}$& [O/H]$_{\rm SME}$ & Error$_{\rm SME}$ & [O/H]$_{\rm EW}$\\
          & (K) &(cm\,s$^{-2}$) & (km\,s$^{-1}$)& &  ($L_{\rm X}/L_{\rm bol}$)$^{b}$  &(mag)&   &\\
\noalign{\smallskip} 
\hline
\noalign{\smallskip}
P\,19  & 6370 & 4.44 & 0.26 & *   & *        &    & $-$0.26 & 0.10 & $-$0.31\\
P\,147 & 6189 & 4.49 & 1.02 & *   & *        &    &    0.08 & *    &    0.00\\
P\,39  & 5867 & 4.50 & 1.25 & *   & *        &0.04& $-$0.03 & *    & $-$0.09\\
P\,107 & 5626 & 4.56 & 0.68 & 5.5 & *        &    &    0.51 & *    &    0.45\\
P\,150 & 5535 & 4.57 & 1.62 & 6.6 & *        &0.06&    0.35 & *    &    0.15\\
P\,151 & 5494 & 4.58 & 1.46 & *   & *        &    &    0.17 & 0.24 &    0.07\\
P\,60  & 5483 & 4.58 & 1.27 & *   & $-$3.44  &    &    0.11 & *    &    0.01\\
P\,75  & 5347 & 4.47 & 1.60 & 6.5 & *        &0.05&    0.46 & *    &    0.19\\
P\,165 & 5292 & 4.59 & 1.56 & 6.6 & *        &    &    0.23 & *    &    0.19\\
P\,267 & 5286 & 4.65 & 0.43 & *   & *        &    & $-$0.03 & *    & $-$0.24\\
P\,64  & 5267 & 4.62 & 0.95 & *   & *        &    &    0.15 & *    &    0.03\\
P\,71  & 5251 & 4.60 & 1.55 & 6.5 & $-$3.11  &0.04&    0.64 & 0.26 &    0.33\\
P\,199 & 5168 & 4.64 & 0.57 & *   & *        &    &    0.22 & *    &    ---$^{d}$\\
P\,94  & 5168 & 4.64 & 0.87 & 6.0 & *        &    & $-$0.20 & *    & $-$0.33\\
P\,100 & 4913 & 4.65 & 1.46 & 6.7 & $-$3.00  &0.10&    1.00 & *    &    0.60\\
\noalign{\smallskip}
\hline
\end{tabular}
\begin{description}
\item $^{a}$ From Mart\'{i}n \& Montes (1997); in units of erg\,cm$^{-2}$\,s$^{-1}$\,\AA$^{-1}$.
\item $^{b}$ From Giampapa, Prosser \& Fleming (1998);
\item $^{c}$ From Allain et al. (1996)
\item $^{d}$ Very low S/N ($<$ 30). No reliable EW measurements could be given for
the shallow oxygen triplet lines.
\end{description}
\end{table}

\clearpage

\begin{table}
\begin{center}
\caption{Oxygen forbidden line abundances of Pleiades dwarfs}
\label{forbid}
\begin{tabular}{lcrcrc}
\hline
\hline
\noalign{\smallskip}
Star & $T_{\rm eff}$ (K) & [O/H]$_{\rm SME}$ & Error$_{\rm SME}$&  [O/H]$_{\rm EW}$ &  Error$_{\rm EW}$\\
\noalign{\smallskip}
\hline
\noalign{\smallskip}
H\,II\,250  & 5731 &    0.02 & 0.08 & $-$0.03 &  0.14\\
H\,II\,129  & 5369 & $-$0.05 & 0.07 & $-$0.02 &  0.14\\
H\,II\,2462 & 5256 & $-$0.10 & 0.11 & $-$0.14 &  0.17\\
H\,II\,522  & 4966 & $-$0.14 & 0.11 & $-$0.17 &  0.18\\
H\,II\,1039 & 4908 &    0.14 & 0.09 &    0.07 &  0.14\\
H\,II\,1298 & 4854 &    0.05 & 0.09 & $-$0.03 &  0.15\\
\noalign{\smallskip}
\hline
\end{tabular}
\end{center}
\end{table}

\begin{table}
\begin{center}
\caption{The surface activity indicators of H$\alpha$ and two \ion{Ca}{2} IRT lines for IC\,4665 stars}
\label{R}
\begin{tabular}{lrrrrrrr}
\hline
\hline
\noalign{\smallskip} 
Star &  R(H$\alpha$) & R(8498) & R(8662) & Err(H$\alpha$) & Err(8498) &Err(8662) & Err(O)\\
\noalign{\smallskip}
\hline
\noalign{\smallskip}
P19  & 0.210 & 0.318 & 0.225 & 0.062 & 0.054 & 0.048 & 0.098\\
P147 & 0.000 & 0.038 & 0.063 & 0.061 & 0.048 & 0.045 & 0.123\\
P39  & 0.274 & 0.350 & 0.303 & 0.045 & 0.040 & 0.039 & 0.168\\
P107 & 0.189 & 0.247 & 0.288 & 0.052 & 0.052 & 0.047 & 0.201\\
P150 & 0.676 & 0.659 & 0.531 & 0.058 & 0.053 & 0.058 & 0.214\\
P151 & 0.341 & 0.535 & 0.482 & 0.062 & 0.056 & 0.051 & 0.219\\
P60  & 0.417 & 0.498 & 0.486 & 0.094 & 0.082 & 0.076 & 0.221\\
P75  & 0.601 & 0.647 & 0.610 & 0.104 & 0.092 & 0.092 & 0.240\\
P165 & 0.851 & 0.637 & 0.511 & 0.090 & 0.079 & 0.090 & 0.247\\
P267 &$-$0.016 &$-$0.103 &$-$0.130 & 0.118 & 0.108 & 0.108 & 0.248\\
P64  &$-$0.086 &$-$0.173 &$-$0.132 & 0.097 & 0.076 & 0.084 & 0.251\\
P71  & 0.731 & 0.666 & 0.613 & 0.160 & 0.143 & 0.127 & 0.253\\
P199 &$-$0.071 &$-$0.109 &$-$0.134 & 0.132 & 0.149 & 0.128 & 0.264\\
P94  & 0.204 & 0.377 & 0.275 & 0.090 & 0.067 & 0.070 & 0.264\\
P100 & 1.186 & 0.951 & 0.733 & 0.173 & 0.127 & 0.132 & 0.300\\

\noalign{\smallskip}
\hline

\end{tabular}
\end{center}
\end{table}

\begin{table}
\begin{center}
\caption{The linear correlation coefficients of oxygen triplet line abundances
with different surface activity indicators}
\label{active}
\begin{tabular}{clllllc}
\hline
\hline
\noalign{\smallskip}
Cluster &  H$\alpha$ & X-ray & \ion{Ca}{2} H+K & \ion{Ca}{2} IRT & A(V)&Ref$^a$\\
\noalign{\smallskip}
\hline
\noalign{\smallskip}
IC\,4665 & 0.34 (7)$^b$    &0.98 (3)   & *       & *         &0.46 (5)& 1, 2, 11\\
         & 0.69 (15)       &           &         & 0.55/0.57$^c$(15)&* &3\\
Pleiades & 0.04 (15)       &0.72 (11)  & *       & 0.64 (15) &* &4, 5, 6\\
M\,34    & 0.15 (8)        &  *        & *       & 0.71 (7)  &* &4, 7\\
UMa      &   *             &  *        &0.77 (6) & *         &* &8\\
Hyades   &   *             &0.16 (28)  &0.21 (40)& *         &* &6, 9, 10\\       
\noalign{\smallskip}
\hline
\end{tabular}
\begin{description}
\item $^{a}$ (1) Mart\'{i}n \& Montes (1997); (2) Giampapa, Prosser \& Fleming
(1998); (3) This work; (4) Schuler et al. (2004); (5) Soderblom et al. (1993); (6) Morel \&
Micela (2004); (7) Soderblom et al. (2001); (8) King \& Schuler (2005), and
references therein; (9) Schuler et al. (2006a); (10) Paulson et al. (2002);
(11) Allain et al. (1996)
\item $^{b}$ Values in parentheses are numbers of stars analyzed.       
\item $^{c}$ For $\lambda$8498 and $\lambda$8662 respectively.
\end{description}
\end{center}
\end{table}

\clearpage

\end{document}